\documentclass[%
reprint,
aps,
pra
]{revtex4-2}

\usepackage[usenames]{color}
\usepackage{colortbl}
\usepackage{todonotes}
\usepackage{graphicx}
\usepackage{dcolumn}
\usepackage{bm}
\usepackage{amsmath}
\usepackage{amssymb}
\usepackage[utf8]{inputenc}

\usepackage{comment}
\usepackage{dsfont}
\usepackage{xcolor}

\begin{document}

\preprint{APS/123-QED}

\title{Kaleidoscopic-ray-tracing-based model of the scintillation flash energy deposition in the photomultipliers attached to a strip scintillator
}

\author{I. V. Vovchenko$^{1}$, T. M. Khamitov$^{1}$, L. V. Inzhechik$^{1}$}
\affiliation{%
 $^1$Moscow Institute of Physics and Technology, 9 Institutskiy pereulok, Dolgoprudny 141700, Moscow region, Russia;
}%


\date{\today}

\begin{abstract}
Strip scintillator detectors are used nowadays in many fields of applied physics, particularly in medicine, civil engineering, mapping of underground resources, and security.
In this work we provide an analytical description of light transport in a cuboid-shaped strip scintillator detector from the scintillation location to the detecting surface.
We use kaleidoscopic-ray-tracing approach to reproduce the average time profile of the energy deposition (light collection) in the detecting surface.
We demonstrate the applicability of the model on the case of PWO crystal of $1.5\times 1.5 \times 30$ cm$^3$ size.
We show that the results achieved in the model are in good agreement with Monte Carlo (MC) simulation.
Notably, the developed model requires dozens of milliseconds to be implemented, thus, it is applicable for real-time calibration of the detector, while the MC simulation takes hours.
Also, we highlight that the kaleidoscopic-ray-tracing itself being used in MC simulations can highly speed them up in the case of specular reflectors.
\end{abstract}

\maketitle








\section{Introduction}
Strip scintillator detectors are used in many areas including high-energy physics~\cite{kobayashi2012performance,chen2007large,korzhik2022ultrafast,lecoq1999scintillator,kharzheev2017scintillation,particle2020review}, medicine~\cite{moskal2016time,moskal2021simulating,casey1997investigation,roncali2011application,evans2006monte,vandenberghe2020state,santarelli2021core,moskal2014test,raczynski2014novel}, civil engineering~\cite{marteau2022development,bonechi2020multidisciplinary,tanaka2023cosmic}, security~\cite{barnes2023cosmic,tanaka2023cosmiccoding,tanaka2023muography,aguiar2015geant4,particle2020review}, geological researches (including natural resources exploration and monitoring)~\cite{tanaka2014particle,tanaka2023muography,holma2022future,zhou2025applications,yin2024simulation,bonneville2017novel,bonneville2018borehole,taira2010potential}, and archaeological researches~\cite{morishima2017discovery,zhou2025muon}.
Strip scintillator detector consist of a long thin cuboid scintillator (organic or inorganic) and a pair of photomultipliers.
It detects high-energy particles passing through the volume of the scintillator, registering the light the particles emit in the scintillation process~\cite{moskal2016time,kobayashi2012performance,marteinsdottir2009light,barnes2023cosmic,birks2013theory,birks1968energy,moszynski1979status}.
The coordinates of the scintillations and the time it has started can be estimated by the analysis of the photomultipliers' output~\cite{moskal2021simulating,moskal2016time,barnes2023cosmic,zhou2025muon,raczynski2014novel}. 
Being arranged in an array, such detectors are able to reconstruct the path of the high-energy particle and its lost energy~\cite{tanaka2014particle,moskal2021simulating,barnes2023cosmic,zhou2025muon,bonneville2017novel}. 

If the flow of high energy particles transits through an object before reaching the detector, the flow's amplitude is reduced~\cite{barnes2023cosmic,tanaka2014particle,zhou2025muon,bonneville2017novel}.
Knowing the geometry of the object one can extract its properties such as density, which could help identifying the object~\cite{marteau2022development,barnes2023cosmic,tanaka2014particle,zhou2025muon,bonneville2017novel,bonechi2020multidisciplinary,pesente2009first,zhou2025muon}.
The objects' sizes can be kilometers as it is in the geology and archaeology researches~\cite{marteau2022development,tanaka2014particle,zhou2025muon,morishima2017discovery}, or dozens of centimeters as it is in security issues~\cite{barnes2023cosmic,pesente2009first}.

The development of a strip scintillator detector starts with the calibration of a single strip~\cite{moskal2016time,moskal2014test,riggi2010geant4,raczynski2014novel,xie2019methods}.
Every longitudinal coordinate of the strip (with some space step) is irradiated with high-energy particles.
As the photons' angle distribution and energy distribution in an individual scintillation event are random, this is repeated multiple times.
For every longitudinal coordinate, the averaged response of the photomultipliers is established.

When a high-energy particle hit the scintillator in use, the response of the photomultipliers is compared to the averaged responses achieved during the calibration.
The most similar averaged response is found, and the longitudinal coordinate of the scintillation is assumed to be coinciding with the longitudinal coordinate of the found most similar average response.
The spatial resolution achieved in this procedure is determined by the randomness of the scintillating process~\cite{donati1969statistical,van2011practical,pizzichemi2016new,moszynski2016energy,xie2019methods,khamitov2024analytical,korevaar2009multi}.

The search of the most similar averaged response involves usage of different metrics and methods to identify the similarities of signals~\cite{moskal2016time,raczynski2014novel,barnes2023cosmic,khamitov2024analytical,korevaar2011maximum,korevaar2009multi,miller2006single,hunter2009calibration,barrett2009maximum,di2024implementing,wang20133d,grinis2022differentiable}.
Usually, analyses of the total energy deposited in the photomultipliers and the analysis of the time profiles are employed.

The calibration process is accompanied with a prior MC simulation~\cite{aguiar2015geant4,bauer2009measurements,rothfuss2004monte,knyazev2021simulations,moskal2016time,riggi2010geant4,xie2019methods}.
The MC modeling delivers the above-mentioned averaged responses of the photomultipliers, allows for estimating the time and spatial resolution of the detector, and can be used to improve the characteristics of the detector by the detector's parameters adjustment in simulation.
Nevertheless, this approach takes a considerable amount of time.
The MC simulation of a detector with a particular set of parameters requires hours and days~\cite{bauer2009measurements,rothfuss2004monte,khamitov2024analytical}. 
Because of that, a real-time calibration of a detector under changing external conditions that affect the geometrical and optical properties of the scintillator (e.g., temperature, pressure, etc.) is hard to conduct.

However, as it is shown in~\cite{khamitov2024analytical,kandarakis2006theoretical,psichis2017analytical}, the average response on the scintillation can be achieved relatively easily in some geometries of the scintillation detector.
Particularly, the detectors consisting of cuboid-shaped scintillators with an attached matrix of photomultipliers allow for such an analytical consideration, delivering the average response of the detector in less than a second of computational time~\cite{khamitov2024analytical}. 
This makes the mentioned analytical approaches usable for real-time calibration of the detectors.

In this work we develop an analytical kaleidoscopic-ray-tracing-based model of light transport in a strip scintillator with attached photomultipliers.
The photons generated in the scintillation flash are tracked by the kaleidoscopic ray-tracing to the detecting surfaces of the photomultipliers.
By that the time profiles of the energy deposited in the detecting surfaces are achieved.
We analyze this deposition of photons in the case of PWO crystal of $1.5\times 1.5 \times 30$ cm$^3$ size.
We show that the results the developed analytical model delivers are in good agreement with MC simulation but require much less time to be achieved, i.e., dozens of milliseconds vs. hours the MC takes.
Also, we highlight that the kaleidoscopic-ray-tracing method itself can highly speed up the MC simulation in the case of specular reflectors, i.e., it allows to consider all the reflections at the lateral faces of the scintillator analytically and exclude them from the simulation.


\section{Kaleidoscopic-ray-tracing-based model of the light transport}

\begin{figure}
    \centering
    \includegraphics[width=\linewidth]{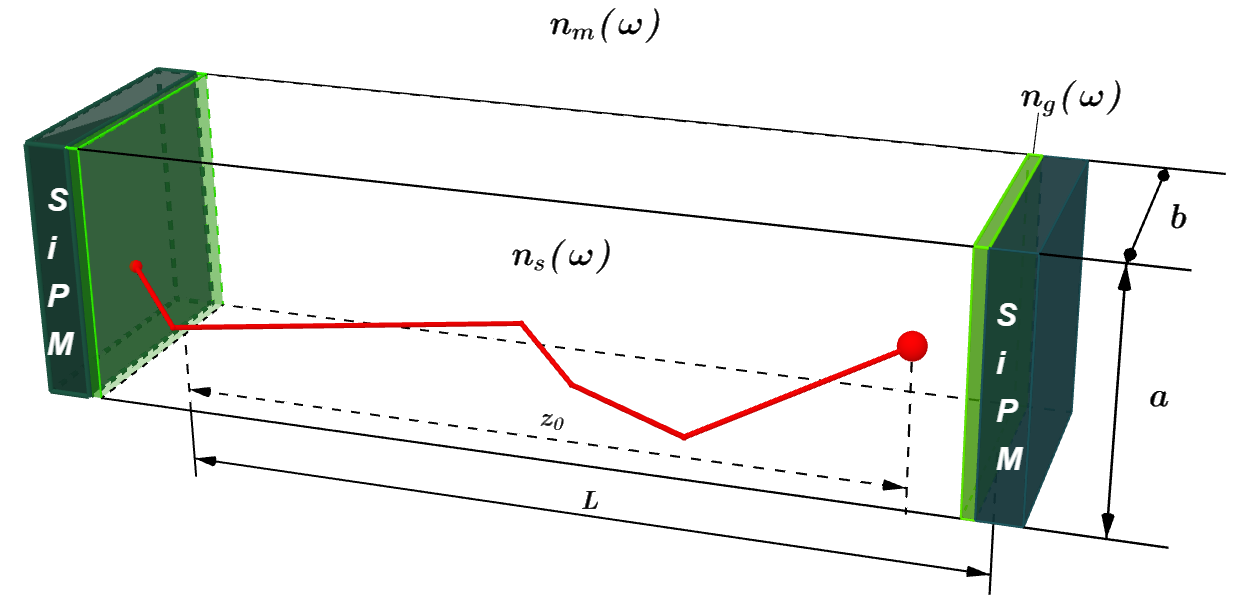}
    \caption{Principal configuration of a strip scintillator detector.
    A long scintillator strip ($L\gg a\sim b$) is attached by its ends to two SiPMs (or other type of photomultiplier) by a layer of optical glue.
    The $n_{s,m,g}(\omega)$ are frequency dependencies of refractive indices of the scintillator strip, the medium it is surrounded by, and the glue ($\omega$ is the frequency of a photon passing through the system).
    For illustration, a local scintillation flash happening at the distance $z_0$ from one of the SiPMs is depicted as a red ball.
    The path of a ray experiencing multiple reflections at the lateral faces of the scintillation strip and reaching this SiPM is depicted.}
    \label{Fig_Detector}
\end{figure}

In this section, we derive a kaleidoscopic-ray-tracing-based model that describes the light transport in a strip scintillator.
We consider a cuboid-shaped strip scintillator of sizes $a\times b \times L$, where $L\gg a\sim b$ (Fig.~\ref{Fig_Detector}).
The photomultipliers are attached to the scintillator faces of $a\times b$ sizes (further, ends of the scintillator).
We denote the refractive index of the scintillator as $n_s(\omega)$, the refractive index of the medium the scintillator is placed in as $n_m(\omega)<n_s(\omega)$, and the refractive index of glue that connects the scintillator ends to photomultipliers as $n_g(\omega)$.
The $\omega$ is the frequency of a photon passing through the scintillator, i.e., we consider the dispersion of the media.

We consider a local scintillation flash of $\gamma$-quantum that happens inside the scintillator far away from its ends.
We denote the time profile of the flash light yeld as $I(t)$.
To describe the collection of the produced photons (rays) having $\omega$ frequency by the photomultiplier, we need to trace them from the flash to the photomultipliers' faces.
We conduct the ray-tracing procedure only for one of the photomultipliers, as for the second one it can be conducted similarly.
(Further, we suppose that photomultipliers are SiPMs.)

When a scintillation flash happens in the scintillator, a part of the yielded light experiences multiple reflections at the scintillator lateral faces (faces with size $a\times L$ and $b\times L$) before reaching a SiPM.
Using the kaleidoscopic ray-tracing technique, we can get rid of the scintillator lateral faces and straighten out these rays, see~\cite{khamitov2024analytical} for details.
After this procedure, the plane that passes through the SiPM's surface would be entirely filled with copies of the SiPM surface.
To describe the collection of the flash's energy by the SiPM, we need to analyze the energy transferred by the photons that are produced in the scintillation flash and pass through this surface~\cite{khamitov2024analytical}.

With this aim, one can mention that some of the straighten rays experience only the total internal reflection during their way towards the SiPM.
For these rays, the scintillator serves as an optical waveguide~\cite{yeh2008essence,marcuse2013theory,jin2008electromagnetic}.
Hence, their intensity is not affected by the reflections at the scintillator lateral faces but is affected by the light attenuation inside the medium~\cite{hara1998doped,aguiar2015geant4,rothfuss2004monte,khamitov2024analytical,mayerhofer2020bouguer,uchida2021attenuation}.
Other rays are reflected at the lateral faces of the scintillator with the reflective coefficient less than $1$~\cite{born2013principles} and are also affected by the light attenuation inside the medium too.

\begin{figure}
    \centering
    \includegraphics[width=0.95\linewidth]{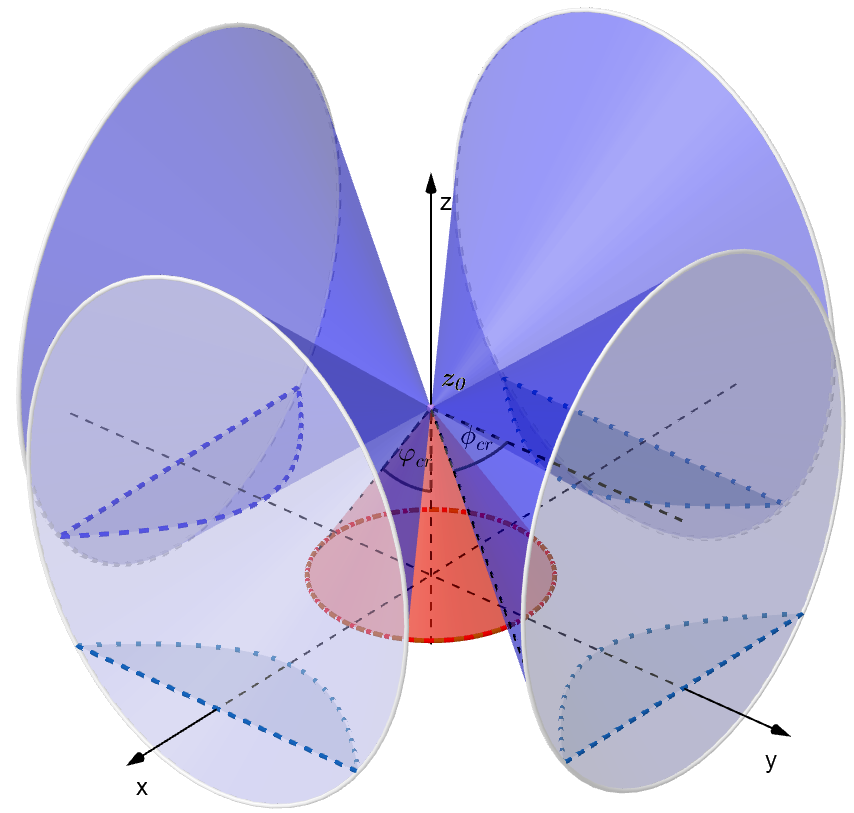}
    \caption{Placement of the critical cones after kaleidoscopic ray-tracing.
    The plane $z=0$ is filled with copies of a SiPM.
    Lateral critical cones draw hyperbolas in the $z=0$ plane, while vertical critical cone draws a circle.}
    \label{Fig_Hyperb_and_circle}
\end{figure}

These two groups of rays are separated by the critical cones directed towards the lateral faces of the scintillator, see blue cones in Fig.~\ref{Fig_Hyperb_and_circle} (further, lateral critical cones).
The first group of rays lies outside the critical cones, while the second group of rays lies inside the critical cones.

The vertex angles of these critical cones equal $2\phi_{cr}(\omega)=2\,{\rm asin}(n_m(\omega)/n_s(\omega))$.
Hence, the rays from the second group experience of the order of $L/ \sqrt{a^2+b^2}\tan \phi_{cr}$ reflections during their way towards the scintillator ends.
Usually, $\tan \phi_{cr}(\omega)\approx 0.5$~\cite{mao2007optical,benaglia2026characterization,bauer2009measurements,kozlova2018optical,khamitov2024analytical,aguiar2015geant4}.
Then, the number of reflections is proportional to $L/\max(a,b)$ that in experiments is usually $10\div 100$~\cite{marteinsdottir2009light,hara1998doped,kobayashi2012performance,zhu2012quality,barnes2023cosmic,bonneville2017novel,bonneville2018borehole}.

The reflection coefficient of the border between two substances with close refractive indices sharply drops down from $1$ to near-zero values when the incidence angle becomes less than the angle of the total internal reflection~\cite{khamitov2024analytical}. 
This perfectly applies to the case of different scintillators immersed in gases or liquids.
Then, even for a reflective coefficient of $0.5$, the output intensity of the rays from the second group is sufficiently diminished, i.e., multiplied by a coefficient sufficiently less than $0.01$.
Thus, further, we neglect the contribution of the rays from the second group to the energy deposited in the SiPM.
(Note that this approximation affects the robustness of the model near the scintillator ends.
It is clear that the length of the volume where the model lacks robustness is of the order of several $\max(a,b)$).

In contrast, the rays that lie outside the lateral critical cones highly contribute to the energy deposited in the SiPM.
Indeed, their intensities are not affected by the reflections at the scintillator lateral faces while the attenuation length varies from dozens of centimeters to several meters~\cite{hara1998doped,aguiar2015geant4,rothfuss2004monte,uchida2021attenuation}.
The lateral critical cones draw hyperbolas in the plane that passes through the SiPM (the conic section~\cite{abbot2013practical,qudosi2021conic}), see Fig.~\ref{Fig_Hyperb_and_circle}.
Hence, a ray can be absorbed by the SiPM if it intersects this plane in-between the hyperbolas.

Additionally, if $n_g(\omega)<n_s(\omega)$, another critical cone is formed by the scintillator-glue border~\cite{khamitov2024analytical}, see red cone Fig.~\ref{Fig_Hyperb_and_circle} (further, vertical critical cone).
The vertex angle of this critical cone equals $2\varphi_{cr}(\omega) =2\,{\rm asin}(n_g(\omega)/n_s(\omega))$.
The cone is directed towards the end of the scintillator.
It draws a circle in the plane that passes through the SiPM.
Thus, in addition to the above, a ray should lie inside the vertical critical cone to be absorbed by the SiPM.

\begin{figure}
\begin{minipage}[h]{0.47\linewidth}
\center{\includegraphics[width=1\linewidth]{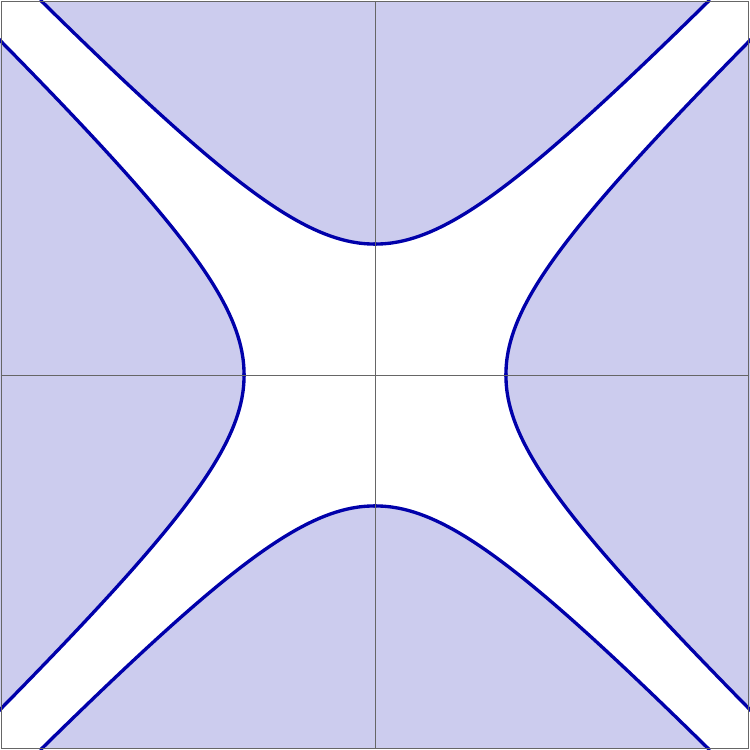}} a) \\
\end{minipage}
\hfill
\begin{minipage}[h]{0.47\linewidth}
\center{\includegraphics[width=1\linewidth]{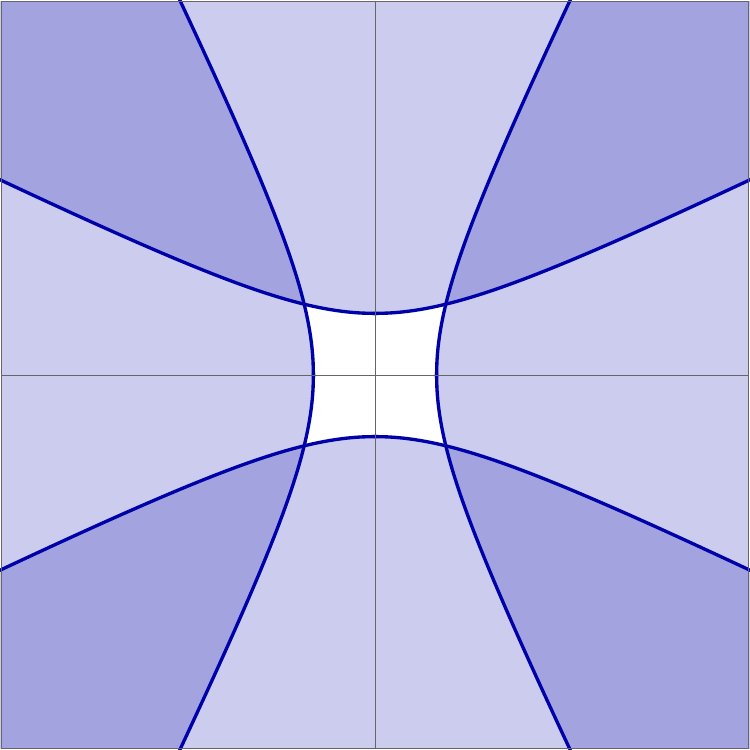}} b) \\
\end{minipage}
\vfill
\begin{minipage}[h]{0.47\linewidth}
\center{\includegraphics[width=1\linewidth]{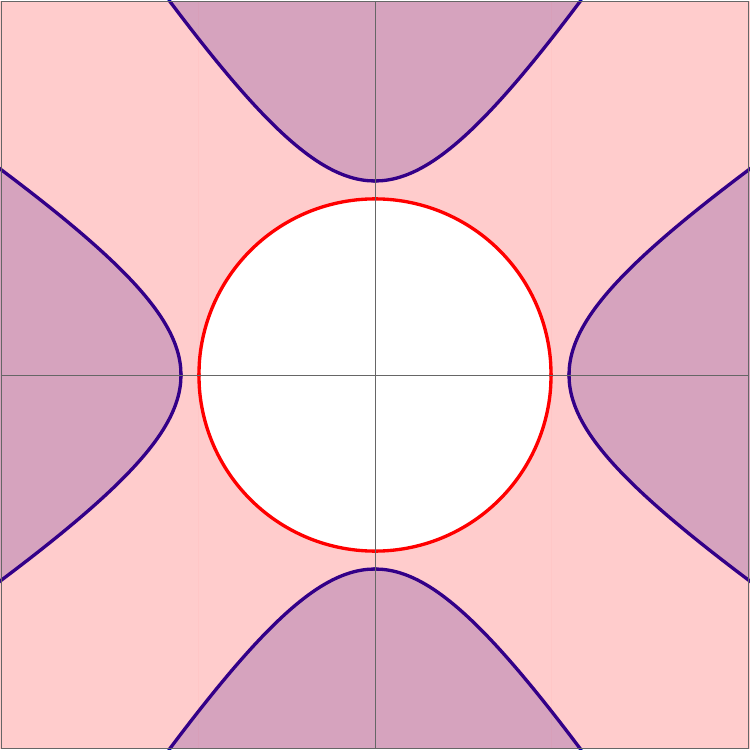}} c) \\
\end{minipage}
\hfill
\begin{minipage}[h]{0.47\linewidth}
\center{\includegraphics[width=1\linewidth]{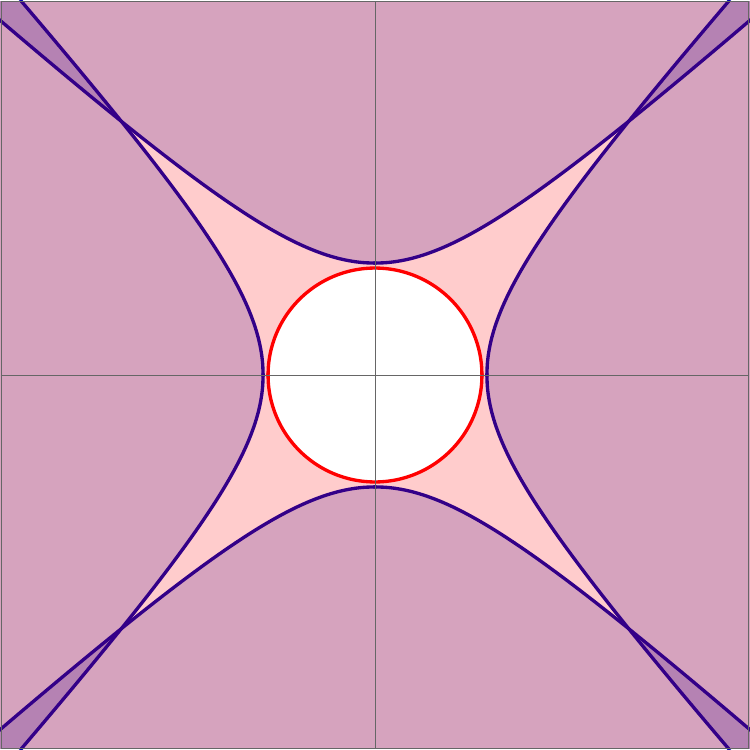}} d) \\
\end{minipage}
\vfill
\begin{minipage}[h]{0.47\linewidth}
\center{\includegraphics[width=1\linewidth]{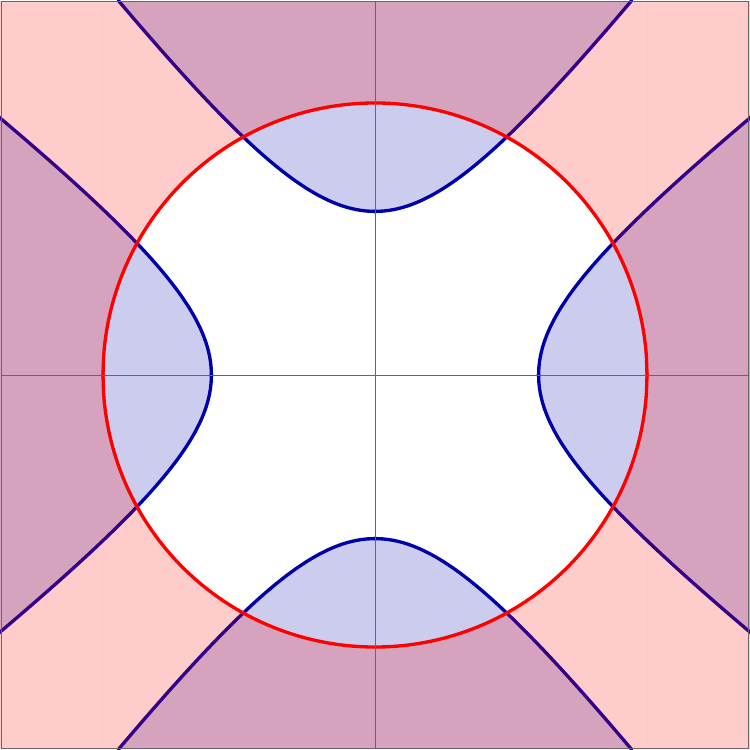}} e) \\
\end{minipage}
\hfill
\begin{minipage}[h]{0.47\linewidth}
\center{\includegraphics[width=1\linewidth]{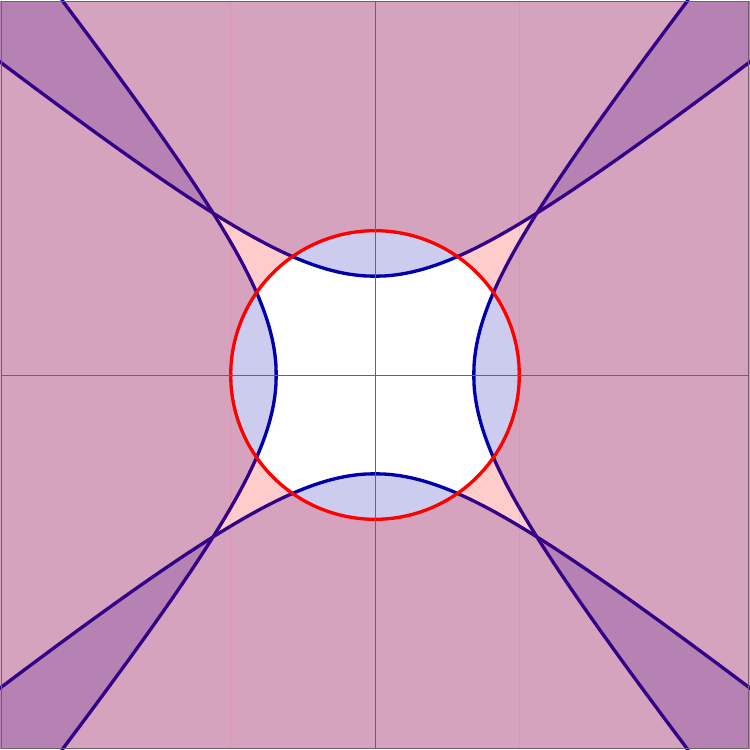}} f) \\
\end{minipage}
\vfill
\begin{minipage}[h]{0.47\linewidth}
\center{\includegraphics[width=1\linewidth]{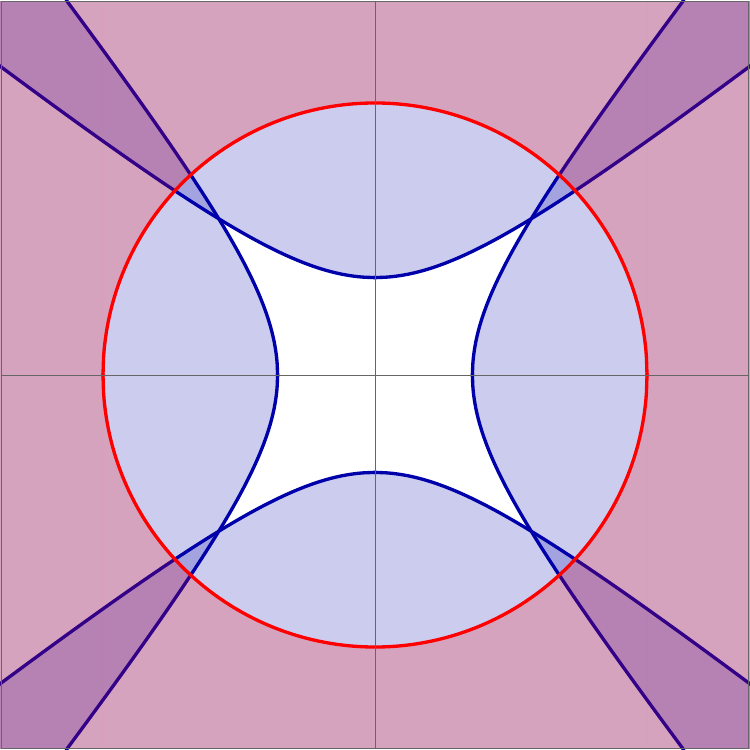}} g) \\
\end{minipage}
\caption{
    The different placement of the hyperbolas (blue curves) and circles (red curves) drawn respectively by the lateral and the vertical critical cones at frequency $\omega$ in the plane that pass through the SiPM in dependence on the ratio of the refractive indexes.
    Areas of absorption are drawn in white.
    The refractive indices are
    (a) $n_s(\omega)=1.46$, $n_m(\omega)=1$, $n_g(\omega)=1.5$, 
    (b) $n_s(\omega)=1.46$, $n_m(\omega)=1.3$, $n_g(\omega)=1.5$,
    (c) $n_s(\omega)=1.85$, $n_m(\omega)=1$, $n_g(\omega)=1.5$,
    (d) $n_s(\omega)=2.15$, $n_m(\omega)=1.6$, $n_g(\omega)=1.4$,
    (e) $n_s(\omega)=1.65$, $n_m(\omega)=1.0$, $n_g(\omega)=1.5$,
    (f) $n_s(\omega)=1.85$, $n_m(\omega)=1.45$, $n_g(\omega)=1.4$,
    (g) $n_s(\omega)=1.65$, $n_m(\omega)=1.3$, $n_g(\omega)=1.5$.}
\label{Hyperb}
\end{figure}

The hyperbolas drawn by the lateral critical cones and the circle drawn by the vertical critical cone can be located differently relative to each other, depending on the ratio of refractive indices $n_s(\omega)$, $n_m(\omega)$, $n_g(\omega)$.
Indeed, the hyperbolas intersect if $\phi_{cr}(\omega)>\pi/4$; if $\phi_{cr}(\omega)+\varphi_{cr}(\omega)>\pi/2$, the hyperbolas intersect with the circle.
Possible variants of the hyperbolas and the circle relative positions are presented in Fig.~\ref{Hyperb}.

The placement in Fig.~\ref{Hyperb}a is typical for low refractive index organic (plastic) scintillators in gas medium, and Fig.~\ref{Hyperb}b is typical for low refractive index organic scintillators immersed in liquid medium.
The placement Fig.~\ref{Hyperb}c is typical for high refractive index inorganic scintillators in gas medium, and Fig.~\ref{Hyperb}d is typical for high refractive index inorganic scintillators immersed in liquid medium.
The placement Fig.~\ref{Hyperb}e is typical for intermediate and high refractive index organic scintillators in gas medium, and Figs.~\ref{Hyperb}f, g are typical for intermediate and high refractive index organic scintillators immersed in liquid medium.
The relative placement of the hyperbolas and the circle defines working area (depicted in white in Fig.~\ref{Hyperb}) of the SiPM's reflections over the lateral surfaces of the scintillator.
This determines the outcome of the SiPM.

The above classification is not a strict rule.
However, it, indeed, describes the frequently occurring ratios of the refractive indices in the experiment.

\begin{figure}
    \begin{minipage}[h]{\linewidth}
    \center{\includegraphics[width=\linewidth]{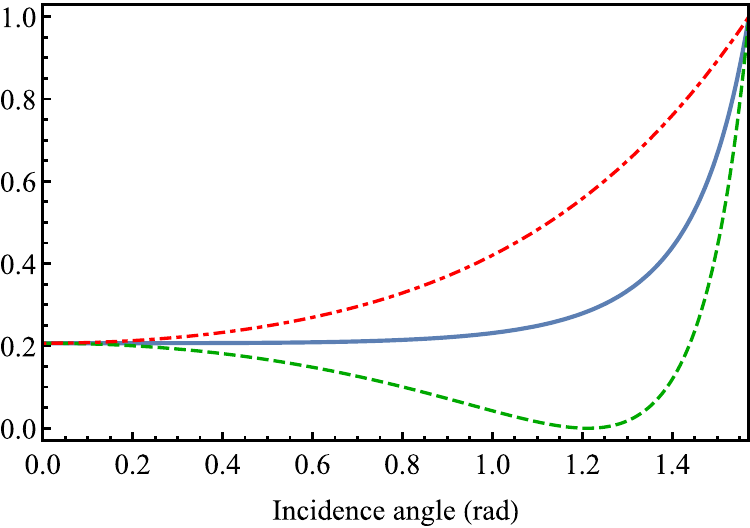}} a) \\
    \end{minipage}
    \vfill
    \begin{minipage}[h]{\linewidth}
    \center{\includegraphics[width=1\linewidth]{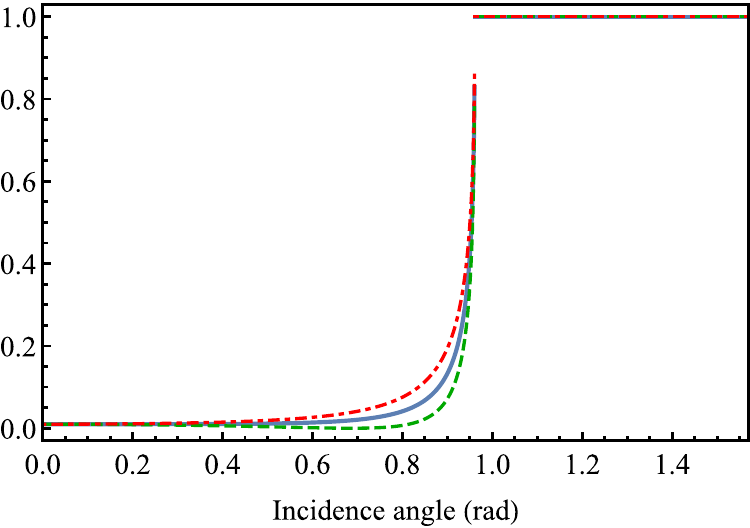}} b) \\
    \end{minipage}
    \caption{The dependencies of (a) glue-SiPM and (b) scintillator-glue borders' intensity reflection coefficients on the incidence angle in natural-polarized light (blue solid line), $p$-polarized light (green dashed line), and $s$-polarized light (red dot-dashed line) according to Fresnel equations~\cite{born2013principles}.
    The SiPM refractive index is set to $4.0$, $n_g=1.5$, $n_s=1.83$.}
    \label{Fig_r_Si_glue}
\end{figure}

Let the light yield time profile of the scintillation flash be equal to $I(t)=\sum_\omega \hbar \omega I_\omega (t)$, where $I_\omega(t)$ is the number of photons with frequency $\omega$ yielded at time $t$ in the scintillation process.
We consider the flash to be isotropic
~\cite{kandarakis2006theoretical,khamitov2024analytical,donati1969statistical,psichis2017analytical,xie2019methods}.

The coordinate system can be introduced such that the scintillation flash would happen at a point with coordinates $(0,0,z_0)$, and the $Oxy$ plane would be the plane that passes through the considering SiPM (see Fig.~\ref{Fig_Hyperb_and_circle}).
First, we can consider the deposition of photons with frequency $\omega$ produced during the time interval $t_s\div t_s+d t_s$ in the SiPM.
We use the polar coordinate system for convenience.
The light from the flash emitted in this time interval reaches the distance $r=\sqrt{x^2+y^2}$ from the point $(0,0,0)$ at time
\begin{gather}
    t=t_s+\frac{\sqrt{r^2+z_0^2}}{c/n_s(\omega)},
\end{gather}
Here $c$ is the speed of light in vacuum.
Note that $t\ge t_s+\tau_0(\omega,z_0)$, where $\tau_0(\omega,z_0)=z_0/(c/n_s(\omega))$ is the minimal time needed for the yielded photons to reach the SiPM.
Hence,
\begin{gather}
    cdt=\frac{n_s(\omega) rdr}{\sqrt{r^2+z_0^2}}.
\end{gather}

The light energy collected in the ring $r\div r+dr$ equals
\begin{gather}
    \frac{I_\omega(t_s)dt_s}{4\pi}
    \frac{z_0 r \Phi(r,z_0,\omega) dr}{(z_0^2+r^2)^{3/2}}
    e^{-\frac{\sqrt{z_0^2+r^2}}{l(\omega)}}.
\end{gather}
Here $l(\omega)$ is the attenuation length at the $\omega$ frequency according to Beer–Bouguer–Lambert law~\cite{mayerhofer2020bouguer}, and $\Phi(r,z_0)$ denotes allowed polar angles according to the working area of the SiPM (see above).
Hence, the light energy reaching the SiPM in $t\div t+dt$ time interval equals
\begin{gather}
    \frac{I_\omega(t_s)dt_s}{4\pi}
    \frac{z_0 \Phi(t-t_s,z_0,\omega) dt}{c(t-t_s)^2/n_s(\omega)}
    e^{-\frac{c(t-t_s)}{n_s(\omega)l(\omega)}}.
\end{gather}

We postulate that $\Phi(t-t_s<\tau_0(\omega,z_0),z_0,\omega)=0$.
Then the average energy absorbed by the SiPM equals
\begin{gather}
    f(t,t_s,\omega,z_0)dt\,dt_s
    =
    {\rm PDE}(\omega)
    {\rm AS}(r(t,t_s),z_0)
    \times
    \\ \nonumber
    \times
    \frac{I_\omega(t_s)dt_s}{4\pi}
    \frac{z_0 \Phi(t-t_s,z_0) dt}{c(t-t_s)^2/n_s(\omega)}
    e^{-\frac{c(t-t_s)}{n_s(\omega)l(\omega)}}.
\end{gather}
Here, ${\rm PDE}(\omega)$ stands for photoelectron detection efficiency, ${\rm AS}(r(t,t_s),\omega)$ stands for angle sensitivity at $\omega$ frequency.
Further, for simplicity, we assume flat ${\rm PDE}(\omega)$~\cite{bonanno2009precision,gallina2019characterization,chen2007large}.
Also, the ${\rm AS}(r(t,t_s),\omega)$ can be considered ${\rm AS}(\omega)$, as the detection surface of the SiPM reflects naturally polarized light with the same reflection coefficient in wide range of incidence angles, see Fig.~\ref{Fig_r_Si_glue}a.
Thus, as it follows from the above equation, ${\rm PDE}(\omega)$ and ${\rm AS}(\omega)$ can be considered by reduction in the number of photons yielded in the scintillation flash.
Further, we denote $D(\omega)={\rm PDE}(\omega){\rm AS}(\omega)$.

To find the time profile of the energy deposition in a SiPM at the moment $t$ we should integrate over $t_s$.
We get
\begin{gather}
    f(t,\omega,z_0)
    \equiv\int\limits_0^{t-\tau_0(\omega,z_0)}dt_s\,f(t,t_s,\omega,z_0)
    =
    \\ \nonumber
    =
    D(\omega)
    \int\limits_0^{t-\tau_0(\omega,z_0)}dt_s\,
        \frac{I_\omega(t_s)}{4\pi}
        \frac{z_0 \Phi(t-t_s,z_0,\omega)}{c(t-t_s)^2/n_s(\omega)}
        e^{-\frac{c(t-t_s)}{n_s(\omega)l(\omega)}}
    \\ \nonumber
    =
    \frac{D(\omega)\tau_0(\omega,z_0)}{4\pi}
    \int\limits_0^{t-\tau_0(\omega,z_0)}
    \hspace{-10pt}dt_s\,
        \frac{I_\omega(t_s) \Phi(t-t_s,z_0,\omega)}{(t-t_s)^2}
        e^{-\frac{(t-t_s)}{\tau_l(\omega)}}
    \\ \nonumber
    =
    \left[\frac{(t-t_s)}{\tau_l(\omega)}\equiv q\right]
    =
    \\ \nonumber
    =
    \frac{D(\omega)\tau_0(\omega)}{4\pi\tau_l(\omega)}
    \int\limits_{\tau_0(\omega,z_0)/\tau_l(\omega)}^{t/\tau_l(\omega)}dq\,
        \frac{\tilde{I}_\omega(q,t) \tilde{\Phi}(q,z_0,\omega)}{q^2}
        e^{-q}.
\end{gather}
Here $\tau_l(\omega)=l(\omega)/(c/n_s(\omega))$ is the time needed for a ray to travel the attenuation length (further, attenuation time), $\tilde{I}_\omega(q,t)$, $\tilde{\Phi}(q,z_0,\omega)$ are functions ${I}_\omega(q)$, ${\Phi}(q,z_0,\omega)$ redefined according to the made substitution.
Finally,
\begin{gather}
    f(t,\omega,z_0)
    \equiv\int\limits_0^{t-\tau_0(\omega,z_0)}dt_s\,f(t,t_s,\omega,z_0)
    =
    \\ \nonumber
    =
    \frac{D(\omega)\eta(\omega,z_0)}{4\pi}
    \int\limits_{\eta(\omega,z_0)}^{\eta(\omega,z_0)t/\tau_0(\omega,z_0)}\hspace{-5pt}dq\,
        \frac{\tilde{I}_\omega(q,t) \tilde{\Phi}(q,z_0,\omega)}{q^2}
        e^{-q}.
\end{gather}
Here $\eta(\omega,z_0)=\tau_0(\omega,z_0)/\tau_l(\omega)$.
Note the existence of two dimensionless parameters in the problem, i.e., $\eta(\omega,z_0)$ and $t/\tau_0(\omega,z_0)$.

We have considered the energy deposited in the SiPM by the photons directed from the flash into the particular hemispace.
However, according to the kaleidoscopic ray-tracing, the reflections of the scintillation flash over the detecting surfaces should be taken into account too~\cite{khamitov2024analytical}.

If $L/l(\omega)\ll 1$, the usual kaleidoscopic ray-tracing procedure is applicable.
If $L/l(\omega)\gg 1$ or $L/l(\omega)\sim 1$, there is no need to consider more than one reflection of the scintillation flash.
Indeed, in Fig.~\ref{Fig_r_Si_glue}a, one can see that the reflective coefficient of the detecting surface in natural light is less than $0.5$ for a wide range of incidence angle.
Hence, multiple reflections over the detecting face induce a drop in the reflected ray intensity.
Additionally, the ray intensity is diminished at least by $e^{-2L/l(\omega)}$ during its travel from one detecting surface to another.
Thus, further, we consider only one reflection of the initial scintillation flash over the detecting surface that is opposite to the considering SiPM.
Then, the time profile of the energy deposition in the SiPM equals
\begin{align}\label{f_tw}
    f(t,\omega)&=\theta(t-\tau_0(\omega,z_0))f(t,\omega,z_0)
    \\ \nonumber
    &+\theta(t-\tau_0(\omega,2L-z_0))rf(t,\omega,2L-z_0).
\end{align}
Here, $\theta(\cdot)$ is the Heaviside step function.
It is included to underline that $t\ge\tau_0(\omega,z_0)$.
The $r$ is SiPM's reflective coefficient at a $0^\circ$ incidence angle.
Here and further, we neglect the dependence of $r$ on the incidence angle, which is sufficient only near $90^\circ$, see Fig.~\ref{Fig_r_Si_glue}a.

Note that the reflections of the rays at the scintillator-glue border do not contribute much to the output intensity, unless it is total internal reflection.
Indeed, in Fig.~\ref{Fig_r_Si_glue}b it is seen that the reflective coefficient of this border is negligible for incidence angles lower than the $\varphi_{cr}(\omega)$.

To find the energy deposition time profile produced by photons of all the frequencies present in the scintillation flash, we just need to sum the Eq.~(\ref{f_tw}) over $\omega$ with weights $\hbar\omega$.
Then, this time profile equals $f(t)=\sum_\omega \hbar \omega f(t,\omega)$.

\subsection{Example of PWO strip crystal}

Here we consider the PWO strip crystal.
The crystal has a stopping power of $0.9$ cm~\cite{mao2007optical} 
and has a good radiation hardness to $\gamma$-quanta~\cite{auffray2016luminescence}.
The peak radiation wavelength is near $420$ nm~\cite{mao2007optical,benaglia2026characterization}, which is close to the maximums of SiPM PDEs~\cite{benaglia2026characterization,roncali2011application,bonanno2009precision,gallina2019characterization,chen2007large}.
The attenuation length of the radiation at the peak wavelength is about $l(\omega)=65$ cm~\cite{hara1998doped}, and the real part of the refractive index equals $n_s(\omega)=2.20$ ~\cite{mao2007optical,benaglia2026characterization}.
Thus, PWO crystal is of interest for applications in high-energy physics and medicine.

The PWO crystals can be grown up to $2\times 2 \times 23$ cm$^3$ size~\cite{marteinsdottir2009light,hara1998doped,ye2006growth,xie2006characterization}.
We test the method developed above on the PWO crystal of $1.5\times 1.5\times 30$ cm$^3$ size.
To analyze the average time profile of the energy deposition in the SiPMs due to the scintillation, we have set the following scintillation flash time profile~\cite{benaglia2026characterization}
\begin{gather}
    I_\omega(t)
    \hspace{-1pt}=\hspace{-1pt}
    0.6\frac{e^{-\frac{t}{3\ {\rm ns}}}\hspace{-1pt}-\hspace{-1pt}e^{-\frac{t}{0.7\ {\rm ns}}}}{2.3\ {\rm ns} }
    \hspace{-2pt}+\hspace{-1pt}0.4\frac{e^{-\frac{t}{10\ {\rm ns}}}\hspace{-1pt}-\hspace{-1pt}e^{-\frac{t}{0.7\ {\rm ns}}}}{9.3\ {\rm ns}}.
\end{gather}
Note that different scintillation time profiles for PWO crystals are reported in literature~\cite{auffray2016luminescence,korzhik2022ultrafast,ye2006growth,auffray2015application,mao2007optical,benaglia2026characterization}.
Here we use a recent result~\cite{benaglia2026characterization}.

Let $n_g(\omega)=1.5$.
Then the placement of the hyperbolas and the circle is depicted in Fig.~\ref{Hyperb}c.
Hence,
\begin{equation}
    \hspace{-2pt}
    \left[
        \begin{array}{l}
            \hspace{-3pt}\Phi(t-t_s, z_0,\omega)\hspace{-2pt}=\hspace{-2pt}0,
            \ \ \ t-t_s<\tau_0(\omega,z_0),\\
            \hspace{-3pt}\Phi(t-t_s, z_0,\omega)\hspace{-2pt}=\hspace{-2pt}2\pi,
            \ \tau_{cr}(\omega,\hspace{-1pt}z_0)\hspace{-2pt}>\hspace{-2pt}t\hspace{-2pt}-\hspace{-2pt}t_s\hspace{-2pt}>\hspace{-2pt}\tau_0(\omega,\hspace{-1pt}z_0),\\
            \hspace{-3pt}\Phi(t-t_s, z_0,\omega)\hspace{-2pt}=\hspace{-2pt}0,
            \ \ \ t-t_s>\tau_{cr}(\omega,z_0),
        \end{array}
    \right.
\end{equation}
Here $\tau_{cr}(\omega,z_0)=z_0\sqrt{1+\tan^2{\varphi_{cr}}}/(c/n_s(\omega))$.
Thus, 
\begin{gather}
    \Phi(t-t_s, z_0,\omega)
    =
    \\ \nonumber
    =
    2\pi(\theta(t-t_s-\tau_0(\omega,z_0))-\theta(t-t_s-\tau_{cr}(\omega,z_0))).
\end{gather}

The following equations are derived for a more general case $I_\omega(t)=\sum_\kappa I_{\omega,\kappa} e^{-t/\tau_{\omega,\kappa}}$ to make them applicable for a wider range of scintillators~\cite{moszynski1979status,birks2013theory,birks1968energy,benaglia2026characterization,mao2007optical,auffray2016luminescence}.
Then,
\begin{gather}
    f(t,\omega,z_0)
    =
    \sum_\kappa
    \frac{I_{\omega,\kappa} D(\omega)\tau_0(\omega,z_0)}{2}
    \times
    \\ \nonumber
    \times\hspace{-13pt}\int\limits_0^{t-\tau_0(\omega,z_0)}\hspace{-15pt}dt_s\,
        \frac{\Phi(t-t_s, z_0,\omega)}{2\pi}
        \frac{
            e^{-t_s/\tau_{\omega,\kappa}}
            e^{-\frac{(t-t_s)}{\tau_l(\omega)}} }
            {(t-t_s)^2}.
\end{gather}

Hence,
\begin{gather}\label{Eq_f_twz0}
    f(t,\omega,z_0)
    =
    \frac{ D(\omega)\tau_0(\omega,z_0)}{2}
    \times
    \\ \nonumber
    \times
    \sum_\kappa I_{\omega,\kappa}
    e^{-t/\tau_{\omega,\kappa}}
    \hspace{-25pt}
    \int\limits_{\max\left(0,t-\tau_{cr}(\omega,z_0)\right)}^{t-\tau_0(\omega,z_0)}\hspace{-15pt}dt_s\,
        \frac{
            e^{\left(
                \frac{1}{\tau_{\omega,\kappa}}
                -\frac{1}{\tau_l(\omega)}
                \right)(t-t_s) }
            }{(t-t_s)^2}
    =
    \\ \nonumber
    =
    \frac{ D(\omega)}{2}
    \sum_\kappa I_{\omega,\kappa}
    e^{-t/\tau_{\omega,\kappa}}
    \alpha_\kappa(\omega,z_0)
    \hspace{-30pt}
    \int
    \limits_{\alpha_\kappa(\omega,z_0)}
        ^{\alpha_\kappa(\omega,z_0)
        \min\left(
            \frac{t}{\tau_0(\omega,z_0)},
            \frac{\tau_{cr}(\omega,z_0)}{\tau_0(\omega,z_0)}
            \right)} \hspace{-30pt}dq
        \frac{e^q}{q^2}.
\end{gather}
Here $\alpha_\kappa (\omega,z_0)=\tau_0(\omega,z_0)/\tau_{\omega,\kappa} - \tau_0(\omega,z_0)/\tau_l(\omega)$.
For positive $\alpha_\kappa (\omega,z_0)$ we use the series instead of the exponent under the above integral
\begin{gather}\label{Eq_Int_e_q_series}
    \int dq\frac{e^q}{q^2}
    =
    \int  \frac{dq}{q^2}
    +\int  \frac{dq}{q}
    +\sum_{n=2}^\infty \int dq \frac{q^{n-2}}{n!}
    =
    \\ \nonumber
    =
    -\frac{1}{q} +  \ln{|q|} + \sum_{n=2}^\infty \frac{q^{n-1}}{n!(n-1)}+{\rm const}.
\end{gather}

Substituting the limits of integration in the last equation, one gets the answer.
The number of terms in the series should be of the order of $2\lceil|\alpha_\kappa(\omega,z_0)|\rceil$ to reproduce the behavior of the exponential function (here $\lceil\cdot\rceil$ is the ceiling operation).

For negative $\alpha_\kappa (\omega,z_0)$ we use the following representation of the integral in Eq.~(\ref{Eq_f_twz0}):
\begin{gather}
    f(t,\omega,z_0)
    =
    \\ \nonumber
    =
    \frac{ D(\omega)}{2}
    \sum_\kappa I_{\omega,\kappa}
    e^{-t/\tau_{\omega,\kappa}}
    |\alpha_\kappa(\omega,z_0)|
    \hspace{-40pt}
    \int
    \limits_{|\alpha_\kappa(\omega,z_0)|}
        ^{|\alpha_\kappa(\omega,z_0)|
        \min\left(
            \frac{t}{\tau_0(\omega,z_0)},
            \frac{\tau_{cr}(\omega,z_0)}{\tau_0(\omega,z_0)}
            \right)} \hspace{-40pt}dq
        \frac{e^{-q}}{q^2},
\end{gather}
where
\begin{gather}
    \int\limits_{q_1}^{q_2} dq\frac{e^{-q}}{q^2}
    =
    \frac{E_2(q_1)}{q_1}-\frac{E_2(q_2)}{q_2},
    \ \ \ \ \ 
    q_2>q_1,
    \\ \nonumber
    E_2(q)
    =
    q\int\limits_q^{+\infty}dq' \frac{e^{-q'}}{q'^2}.
\end{gather}
Here $E_2(q)$ is the generalized exponential integral of index $2$.
This representation is reliable for large negative powers of the exponent, where the series representation reveals sufficient numerical errors~\cite{progri2022study}.

For the intermediate case $\alpha_\kappa (\omega,z_0)=0$, i.e., when $\tau_{\omega,\kappa}=\tau_l(\omega)$, the integral in Eq.~(\ref{Eq_f_twz0}) can be easily computed
\begin{gather}
    f(t,\omega,z_0)
    =
    \frac{ D(\omega)\tau_0(\omega,z_0)}{2}
    \times
    \\ \nonumber
    \times
    \sum_\kappa I_{\omega,\kappa}
    e^{-t/\tau_{\omega,\kappa}}
    \hspace{-25pt}
    \int\limits_{\max\left(0,t-\tau_{cr}(\omega,z_0)\right)}^{t-\tau_0(\omega,z_0)}\hspace{-15pt}dt_s\,
        \frac{ 1}{(t-t_s)^2}
    =
    \\ \nonumber
    =
    \frac{ D(\omega)}{2}
    \sum_\kappa I_{\omega,\kappa}
    e^{-\frac{t}{\tau_{\omega,\kappa}}}\hspace{-2pt}
    \left(
        1\hspace{-2pt}-\hspace{-2pt}\frac{\tau_0(\omega,z_0)}{t-\max(0,t-\tau_{cr}(\omega,z_0))}
    \right)\hspace{-2pt}.
\end{gather}

\section{MC simulation}

We have conducted an MC simulation of light transport in the PWO crystal analyzed above to check Eq.~(\ref{Eq_f_twz0}).
In general, MC of a scintillation detector involves three main layers.
Each layer simulates particular physical processes.
The first layer simulates the light emission in the scintillation.
The second layer simulates light transport from the location of the scintillation to the photomultipliers.
The third level simulates the response of the photomultipliers to the incident light. 
Particularly, the Geant4 simulation toolkit \cite{geant4} employs the first two layers of this scheme. The third layer can be employed by \cite{sipm_lib}, which simulates the SiPM's response to the absorption of the incident photons.

The second layer that is responsible for the simulation of the light transport interacts with the other layers through the creation and deletion of photons. 
In this paper we focus on the light transport in strip scintillator.
Hence, the scintillation physics could be omitted in MC, assuming an isotropic scintillation flash~\cite{kandarakis2006theoretical,khamitov2024analytical,donati1969statistical,psichis2017analytical,xie2019methods}.
Also, we neglect the Poison statistics of photon number counting in the SiPMs~\cite{Loudon}, considering only the deposition of the light energy in the detecting surfaces.
In other words, we employ only the second layer of an usual MC simulation in our MC to verify the results of the analytical approach we have developed
(the developed approach only describes the light transport in the scintillator).

We choose the pvtrace \cite{pvtrace_raytracing} package to track photons.
Pvtrace was verified by its author via comparison with a thermodynamic model and two other ray tracers. 
Pvtrace has a sufficient drawback when time profiles are studied.
It does not count the photon's travel time, so a fork \cite{pvtraceFork} that adds this feature was created. 

We register all the photons in the detector volume independent of their energy, but accounting for the reflection of the detecting surface according to Fresnel equations~\cite{born2013principles}. 

The photon emission is sampled from the scintillation time distribution with a $0.1$ ns binning step.
We have simulated average time profiles of the energy deposited in the detecting surfaces of the SiPMs for scintillation flashes happening at $0$ cm, $3$ cm, $6$ cm, $9$ cm, and $12$ cm away from the center of the crystal along its longest side. 
In each coordinate $50$ events were simulated (the event is an isotropic light yield from the particular point in the scintillator volume). 
The number of the generated photons in each event follows the Poisson distribution with expected value $20000$.
The event coordinates in the cross-section plane perpendicular to the longest side of the crystal were sampled from the two-dimensional uniform distribution.

\begin{figure}
    \begin{minipage}[h]{\linewidth}
    \center{\includegraphics[width=\linewidth]{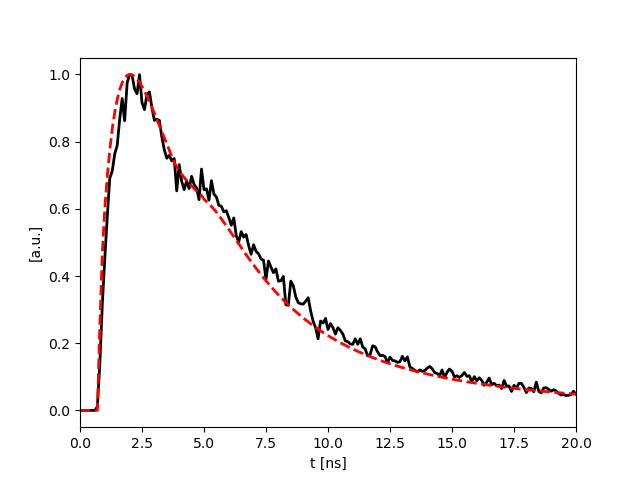}} a) \\
    \end{minipage}
    \vfill
    \begin{minipage}[h]{\linewidth}
    \center{\includegraphics[width=1\linewidth]{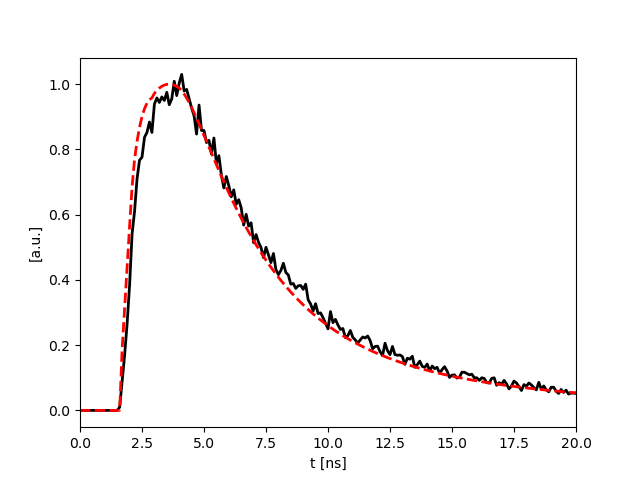}} b) \\
    \end{minipage}
    \caption{The energy deposition in the detecting surfaces of (a) the closest and (b) the furthest SiPM in MC (black solid line) and according to Eq.~(\ref{Eq_f_twz0}) (red dashed line) for  PWO strip crystal of $1.5\times 1.5\times 30$ cm$^3$ size.
    The coordinate of the scintillation flash is $6$ cm away from the crystal center along $z$-axis.
    The SiPM refractive index is set to $4.9$, $n_g=1.5$, $n_s=2.2$, $n_m=1.0$.}
    \label{Fig_SF_21}
\end{figure}

Scintillator crystal absorption is considered by the pvtrace with a $65$ cm attenuation length following the Beer–Bouguer–Lambert law~\cite{mayerhofer2020bouguer}.
All generated photons are tracked until they reach one of the following states: (i) the photon escapes the world's volume, which contains all other volumes, (ii) the photon is absorbed in the material, (iii) the photon penetrates the SiPM's volumes, i.e., it is detected.
For each photon that has reached the (iii) state, we store information about which SiPM was hit and the moment in time this has happened.

\begin{figure}
    \begin{minipage}[h]{\linewidth}
    \center{\includegraphics[width=\linewidth]{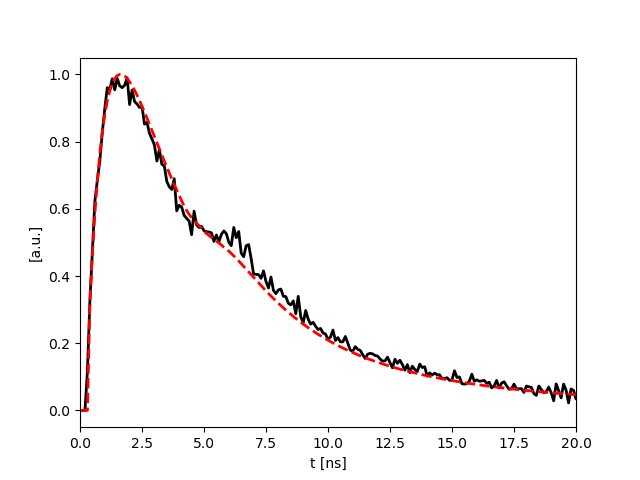}} a) \\
    \end{minipage}
    \vfill
    \begin{minipage}[h]{\linewidth}
    \center{\includegraphics[width=1\linewidth]{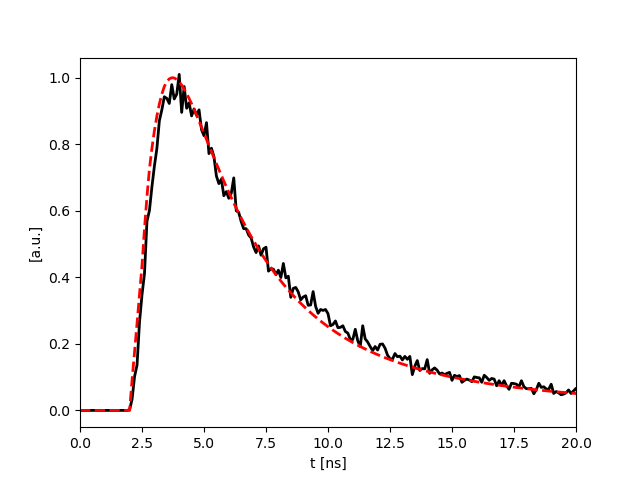}} b) \\
    \end{minipage}
    \caption{The same dependencies as in Fig.~\ref{Fig_SF_21} but the coordinate of the scintillation flash is $12$ cm away from the crystal center along $z$-axis.}
    \label{Fig_SF_27}
\end{figure}

Several MC time profiles of the energy deposition in the SiPMs are presented in Fig.~\ref{Fig_SF_21}-\ref{Fig_SF_27}. 
It is seen that the analytical averaged energy deposition time profiles achieved by Eq.~(\ref{Eq_f_twz0}) approximate the average MC time profiles.
Particularly, it is seen that the highest peaks and the arrival of the first photons to the detectors are reproduced quite precisely in the analytical model as well as the energy deposition tails.

The analytical model also reproduces the additional broadening of the energy deposition caused by the reflection of the photons from the opposite SiPM.
Indeed, the overlay of this signal to the main signal produced by the photons targeting the SiPM directly can be easily seen in MC in Fig.~\ref{Fig_SF_21}a, \ref{Fig_SF_27}a.
This effect creates short shelves in the energy deposition time profiles after the highest peak and broadens the time profile.
The analytical approach predicts smoother shelves and slightly narrower broadening of the time profiles, revealing these two features qualitatively.

Note that, unexpectedly, the analytical model also works well near the end faces of the scintillator crystal where the accomplishment of the assumptions made for the simplification of the analytical description, see above, is questionable. 

\begin{figure}
    \centering
    \includegraphics[width=\linewidth]{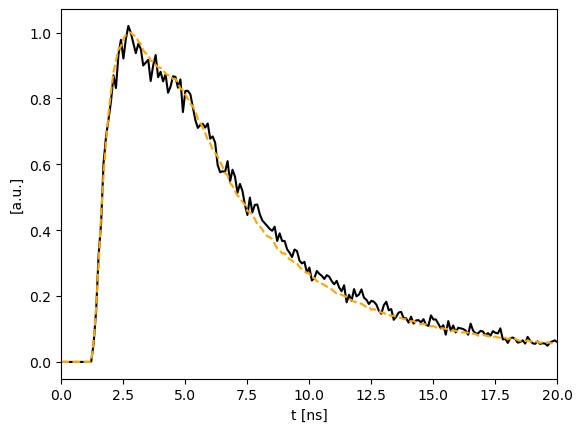}
    \caption{The energy deposition in the detecting surface of a SiPM in MC (black solid line) and MC with specular reflections at lateral faces of the scintillator being excluded via kaleidoscopic ray-tracing (orange dashed line).
    The PWO strip crystal of $1.5\times 1.5\times 30$ cm$^3$ size is considered.
    The coordinate of the scintillation flash is $0$ cm away from the crystal center along $z$-axis.
    The SiPM refractive index is set to $4.9$, $n_g=1.5$, $n_s=2.2$, $n_m=1.0$.}
    \label{Fig_MC_KR}
\end{figure}

Thus, the developed analytical model reproduces the main features of the energy deposition time profiles that are analyzed in the experiment to calculate the spatial resolution of the detector~\cite{moskal2016time,vandenberghe2020state,raczynski2014novel}.
Notably, the analytical time profile can be achieved in $30$ ms, that is quite fast compared to several hours the MC takes.
Therefore, the developed analytical model can be utilized for real-time calibration of the scintillator detectors.

Also note that the kaleidoscopic ray-tracing can be involved directly in the MC simulation.
This can sufficiently reduce the computational time, as all the specular reflections at the lateral faces of the scintillator can be excluded from the simulation.

Indeed, after the kaleidoscopic reflection is done, one only needs to check if the trajectory of the created photon intersects the white area in Fig.~\ref{Hyperb}.
An easy way to do this is to compute angles between the $x,y,z$-axes in Fig.~\ref{Fig_Hyperb_and_circle} and the trajectory of the photon.
After that one can compare these angles with $\phi_{cr}(\omega)$ and $\varphi_{cr}(\omega)$ and conclude if the photon's trajectory intersects the mentioned above white area.

This composite approach sufficiently simplifies the second layer of the MC simulation, lowering the number of computations that should be conducted.
The comparison between this variation of the MC and the usual MC discussed above is represented in Fig.~\ref{Fig_MC_KR}.
It is seen that the MC employing the kaleidoscopic ray-tracing delivers a result similar to the usual MC.
Notably, the computational time in the case of the MC employing the kaleidoscopic ray-tracing was several seconds while the usual MC has taken several hours.

\section{Discussion and conclusion}

In this paper we have presented an analytical approach to the description of the light transport in the strip scintillator with photomultipliers attached to its end faces.
The approach is based on the kaleidoscopic ray-tracing.
We have investigated the average time profiles of energy deposition in detecting surfaces of the photomultipliers.
The equations describing these time profiles and their dependencies on the parameters of the problem, such as the distance between the scintillation flash and the detecting surface, the refractive indices of the substances, the attenuation length in the scintillator, and the scintillation flash time profile, were derived.

We have tested the suggested method on the example of a $1.5\times 1.5 \times 30$ cm$^3$ size PWO crystal.
It is shown that the MC simulation delivers close results to the developed method.
Nevertheless, the developed method requires dozens of milliseconds to be conducted, while the MC simulation has taken several hours.
Thus, the developed method can be used for real-time calibration of the scintillator detectors in continuously changing external conditions, such as temperature and pressure.

Besides that, we note that the kaleidoscopic ray-tracing can be involved in MC simulation directly.
This can sufficiently decrease the computational time, as all the specular reflections at the lateral faces of the scintillator can be considered analytically and dropped out from the simulation.


\bibliography{Scint}

\end{document}